\documentclass[prd,preprint,showpacs,preprintnumbers]{revtex4}
\usepackage{amsmath} 
\usepackage{amsfonts} 
\usepackage{amssymb}
\usepackage{graphicx}

\begin{document}

\preprint{MIT-CTP-3730}
\preprint{KEK-TH-1077}
\preprint{OIQP-06-02}

\title{
Anomalies, Hawking Radiations and Regularity \\
in Rotating Black Holes
} 
\author{Satoshi Iso$^{a}$} 
\email{satoshi.iso@kek.jp}
\author{Hiroshi Umetsu$^b$}
\email{hiroshi_umetsu@pref.okayama.jp}
\author{Frank Wilczek$^c$}
\email{wilczek@mit.edu}  

\affiliation{
$^a$ Institute of Particle and Nuclear Studies, High Energy Accelerator
Research Organization(KEK),  
Oho 1-1, Tsukuba, Ibaraki 305-0801, Japan \\
$^b$ Okayama Institute for Quantum Physics,
Kyoyama 1-9-1, Okayama 700-0015, Japan\\
$^c$ Center for Theoretical Physics, Laboratory for Nuclear
Science and Department of Physics, Massachusetts Institute of Technology,
Cambridge, Massachusetts 02139, USA 
}

\begin{abstract} 

This is an extended version of our previous 
letter \cite{Iso:2006wa}.  In this paper we consider rotating black holes
and show that the flux of Hawking radiation 
can be determined by
anomaly cancellation conditions and regularity
requirement at the horizon.
By using a dimensional reduction technique,  each partial wave of
quantum fields in  a $d=4$ rotating black hole background can be interpreted  
as a $(1+1)$-dimensional charged field with a charge proportional to
the azimuthal angular momentum $m$.
From this and the analysis \cite{Robinson:2005pd} \cite{Iso:2006wa} on
Hawking radiation from charged black holes, 
we show that the total flux of Hawking radiation from rotating black holes
can be universally determined in terms of the values of anomalies
at the horizon by demanding  
gauge invariance and  general coordinate covariance at the quantum level. 
We also clarify our choice of boundary conditions 
and show that our results are consistent 
with the effective action approach where regularity at the 
future horizon and vanishing of ingoing modes at $r=\infty$
are imposed (i.e. Unruh vacuum). 

\end{abstract}

\maketitle

 
\section{Introduction}
Hawking radiation is the most prominent quantum effect 
to arise for quantum fields in a background space-time with an event
horizon. There are several derivations and all of them take the quantum
effect in black hole backgrounds into account in various ways. 
The original derivation by Hawking \cite{Hawking:sw} \cite{Hawking:rv} 
calculates the Bogoliubov coefficients
between the in and out states in a black hole background. 
A tunneling picture \cite{Parikh:1999mf,tunneling}
is based on pair creations of particles and antiparticles near the horizon
and calculates WKB amplitudes for classically forbidden paths. 
A common property in these derivations is the universality of the radiation: 
i.e. Hawking radiation is determined universally by the horizon properties
(if we neglect the grey body factor induced by the effect of scattering
outside the horizon.) 

Another approach to the Hawking radiation is to calculate the
energy-momentum (EM) tensor in the black hole backgrounds.
It has a long history and there are many 
investigations 
(see for example \cite{Frolov:1998wf} and references therein). 
Here we would like to
mention the seminal work by Christensen and Fulling
\cite{Christensen:jc}. In this paper the authors determined the
form of the EM tensor by using 
symmetry arguments and the conservation law
of the EM tensor together with the trace anomaly.
In $d=2$, such information is sufficient to
determine the complete form of the EM tensor and accordingly the Hawking
radiation can be correctly reproduced.
But in $d=4$ there remains
an indeterminable function and the full form of the EM
tensor can not be determined by symmetries only.
Since the Hawking radiation is a very universal phenomenon, 
it should be discussed based on fundamental properties at the horizon.

In a previous paper \cite{Iso:2006wa}, 
we have shown that the flux of Hawking radiation from 
Reissner-Nordstr\"om black holes can be 
determined by requiring gauge and general coordinate covariance at the
quantum level. 
The work was based on \cite{Robinson:2005pd} but with a slightly different
procedure.  
In the following we take the procedure adopted in \cite{Iso:2006wa}.
The basic idea  is the following.
We consider a quantum field in a black hole background.
Near the horizon, the field can be effectively described by
an infinite collection of (1+1)-dimensional fields on $(t,r)$ space
where $r$ is the radial direction.  Furthermore, due to the property of the
black hole metric, mass or potential terms for quantum fields in it 
can be suppressed near the horizon.
Therefore we can treat the original higher dimensional
theories as a collection of  two-dimensional quantum fields. 
In this two-dimensional reduction, 
outgoing modes near the horizon behave as 
right moving modes while ingoing modes as
left moving modes. 
Since the horizon is a null hypersurface, 
all ingoing modes at the horizon can not 
{\it classically} affect physics outside the horizon.
Then, if we integrate the other modes to obtain the 
effective action in the exterior region, it becomes
anomalous with respect to gauge or general 
coordinate symmetries since the effective
theory is  now chiral at the horizon. 
The underlying theory is of course invariant under these symmetries
and these anomalies must be cancelled by quantum effects
of the classically irrelevant ingoing modes. 
We have shown that the condition for anomaly cancellation
at the horizon  determines the Hawking flux of the charge and
energy-momentum.  The flux is universally determined only by
the value of anomalies at the horizon. 

In this paper, we further extend the analysis to Hawking radiations
of quantum fields from rotating black holes. 
In the case of Kerr black hole, the metric is axisymmetric and the azimuthal
angular momentum is conserved. Because of this isometry, the effective
two-dimensional theory for each partial wave
has $U(1)$ gauge  symmetry. 
The effective background gauge potential
for this $U(1)$ symmetry is written 
in terms of the metric while the quantum field 
in the Kerr background has a charge $m$ 
of this gauge symmetry, where $m$ is an azimuthal
quantum number. 
The effective theory is now interpreted as a two-dimensional field theory 
of charged particles in a charged black hole.
Hence we can apply the same 
method for the charged black holes to 
obtain the Hawking flux from rotating black holes. 

Our calculation based on anomaly cancellations 
reproduces the Hawking fluxes in the so-called Unruh vacuum
\cite{Unruh:db}.  
This vacuum violates the time reversal symmetry
by boundary conditions. Namely regularity at the future
horizon is imposed, which fixes the flux of 
the outgoing modes. On the other hand, for ingoing modes,
it is assumed that there is no ingoing flux at $r=\infty.$
In deriving the flux in our method, we demand that
covariant current at the horizon should vanish. We show that
this is nothing but the regularity condition.
On the other hand, the boundary condition for ingoing
modes is also imposed. We clarify these
points in this paper.

The content of the paper is as follows.
In section \ref{QFK}, we will first show that,
near the horizon of Kerr black hole,
each partial wave of  scalar fields  
behaves as a charged field in two-dimensional
charged black hole. In section \ref{SC}, we relate
symmetries in the original $d=4$ system and
the dimensionally reduced (1+1)-dimensional system
and derive modified conservation laws of current
and energy-momentum tensor in $d=2$. 
By using the dimensional reduction technique, we
derive the Hawking flux from Kerr black hole in section 
\ref{AHF}. Here we demand gauge and general coordinate
covariance at the horizon and impose that the covariant
currents should vanish at the horizon.
We discuss our choice of boundary conditions.
In section \ref{KN} we  derive the flux of Hawking radiation from
Kerr-Newman black holes. 
Section \ref{CD} is devoted to conclusions and discussions. 
In Appendix \ref{A1} we calculate the flux of Hawking radiation
from Kerr-Newman black hole by integrating the Planck distribution. 
In Appendix \ref{A2} we derive the flux of radiation from charged 
black holes by an effective action approach. 
Since quantum fields near horizons of black holes 
are effectively described by (1+1)-dimensional conformal fields,
we can explicitly calculate expectation values of 
a current or an energy-momentum tensor near the horizon
by imposing boundary conditions, such that physical
quantities are regular at the future horizon and there is no
ingoing flux at $r=\infty.$ This is the boundary condition
for the Unruh vacuum and the fluxes coincide with those
derived in this paper. 
In Appendix \ref{A3} we summarize our notations for the 
Kaluza-Klein reduction to clarify the conservation equations
of the energy-momentum tensor in $d=2.$

\section{Quantum Fields in Kerr Black Hole \label{QFK}}
The metric of the Kerr black hole is given by
\begin{eqnarray}
 \label{Kerrmetric}
 ds^2 &=& \frac{\Delta}{r^2+a^2\cos^2\theta}
  \left(dt - a\sin^2\theta d\varphi\right)^2
  -\frac{\sin^2\theta}{r^2+a^2\cos^2\theta}
  \left(adt - (r^2+a^2)d\varphi\right)^2 \nonumber \\
 && -\left(r^2+a^2\cos^2\theta\right)
  \left(\frac{dr^2}{\Delta} + d\theta^2\right),
\end{eqnarray}
where
\begin{equation}
 \Delta = r^2 - 2Mr + a^2 = (r-r_+)(r-r_-),
 \end{equation}
and
$r_{+(-)}$ are radii of outer (inner) horizons
\begin{equation}
r_\pm = M \pm \sqrt{M^2-a^2}.
\end{equation}

We will consider matter fields in the Kerr black hole
background. For a while we will consider a scalar field
for simplicity. The action consists of the free part 
\begin{eqnarray}
 S_{free} &=& \frac{1}{2}\int d^4x \sqrt{-g} 
  ~g^{\mu\nu}\partial_\mu\phi \partial_\nu
  \phi \nonumber \\
 &=& -\frac{1}{2}\int dtdrd\theta d\varphi \sin\theta ~\phi
  \left[
   \left(\frac{(r^2+a^2)^2}{\Delta}-a^2\sin^2\theta\right)\partial_t^2
   +2a\left(\frac{r^2+a^2}{\Delta}-1\right)\partial_t \partial_\varphi 
   \right. \nonumber \\
 && \hspace{20mm} 
  \left.
   -\partial_r\Delta\partial_r
   -\frac{1}{\sin^2\theta}\partial_\theta \sin^2\theta\partial_\theta
   -\frac{1}{\sin\theta^2}
   \left(1-\frac{a^2\sin^2\theta}{\Delta}\right)\partial_\varphi^2
  \right] \phi,
\end{eqnarray}
and the other parts $S_{int}$ including  a mass term, potential terms
and interaction terms.
Performing the partial wave decomposition of $\phi$ in terms of the
spherical harmonics, $\phi = \sum_{l,m} \phi_{lm} Y_{l,m}$,
the theory is reduced to a two-dimensional effective theory
with an infinite collection of fields with quantum numbers
$(l,m)$.  These two-dimensional fields are interacting 
with each other. Because of the axial symmetry
of the Kerr black hole metric in the $\varphi$ direction, 
the azimuthal quantum number, $m$ of $Y_{lm}$, is
conserved.

Upon transforming to the $r_*$ tortoise coordinate, defined by 
\begin{equation}
 \frac{dr_*}{dr} = \frac{r^2+a^2}{\Delta} \equiv f(r)^{-1},
\end{equation}
and considering the region near the outer horizon $r_+$, 
one finds that the effective two-dimensional 
action is  much simplified.
The effective radial potentials for partial waves 
($ \sim l(l+1)/r^2)$
or mixing terms between fields with different angular momenta contain a
suppression factor 
$f(r(r_*))$ and vanish exponentially fast near the horizon.
The same applies to a mass term or interaction terms 
$S_{int}$.
Thus it is straightforward to show that the physics near the horizon can be
effectively described by an infinite collection
of massless $(1+1)$-dimensional fields with the following action, 
\begin{eqnarray}
 \label{2dim-Seff}
 S = \int dtdr ~(r^2+a^2) \phi_{lm}^*
  \left[
   \frac{r^2+a^2}{\Delta}
   \left(\partial_t+\frac{iam}{r^2+a^2}\right)^2
   -\partial_r\frac{\Delta}{r^2+a^2}\partial_r
  \right] \phi_{lm}.
\end{eqnarray}

From this action we find that $\phi_{lm}$ can be 
considered as  a $(1+1)$-dimensional complex scalar field in the
backgrounds of the dilaton $\Phi$, metric $g_{\mu\nu}$ and $U(1)$ gauge
field $A_\mu$,
\begin{eqnarray}
 \label{2d-bg}
 && \Phi = r^2 + a^2, \nonumber \\
 && g_{tt} = f(r), \qquad 
  g_{rr} = -\frac{1}{f(r)}, \qquad
  g_{rt} = 0,\\
 && A_t = -\frac{a}{r^2 + a^2}, \qquad
  A_r = 0. \nonumber
\end{eqnarray} 
The $U(1)$ charge of the two-dimensional field 
$\phi_{lm} $ is $m$.


\section{Symmetries and Conservation Laws \label{SC}}
 
The $U(1)$ gauge symmetry in the effective two-dimensional theories
originates from the axial isometry of the Kerr black hole. 
Since the fields are in the background of 
dilaton and gauge potentials, the conservation
law for the energy-momentum tensor is modified.
 
In this section we will see how the $U(1)$ symmetry
arises from the general coordinate invariance in the axial direction
and then show that the conservation law for the energy-momentum tensor in
$d=4$ is rewritten as modified conservation laws of the $U(1)$ current and  
energy-momentum tensor in $d=2$.

The quantum fields in the $d=4$ Kerr black hole background
is invariant under the general coordinate symmetries. 
In particular we are interested in the general coordinate transformations
in the $\varphi$ direction $\xi^\varphi$ which is 
independent of the angles $(\theta, \varphi)$. 
They generate the $U(1)$ 
gauge transformations with
 the transformation parameter 
$\xi^\varphi(t,r)$.
For such transformations, since the metric transforms as 
$\delta g^{\mu \nu}=-(\nabla^\mu \xi^\nu + \nabla^\nu \xi^\mu)$, 
we can define a gauge potential
as $A^{\mu}=- g^{\mu \varphi}$ with a transformation
$\delta A^\mu= \nabla^\mu \xi^\varphi$. Here 
$\mu$ is restricted to $t$  or $r$.  $g^{\varphi \varphi}$
is interpreted as a dilaton which is invariant under the above
transformation. 
A matter field $\phi_{lm}$ transforms as a charged
field with a charge $m$; $\delta \phi_{lm}=im \xi^\phi \phi_{lm}$.

The energy-momentum tensor $T^{\mu\nu}$ of matter fields
in the Kerr black hole background is conserved in $d=4,$
\begin{equation}
 \label{EM-cons}
  \nabla_\nu T^{\mu\nu} = 0.
\end{equation}
Since the Kerr background is 
stationary and axisymmetric, 
the expectation value of the energy-momentum tensor in the background 
 depends only on $r$ and $\theta$, i.e.  
$\langle T^{\mu\nu} \rangle =\langle T^{\mu\nu}(r, \theta)\rangle$. 
(In the following we omit the bracket for notational simplicity. )

First, the $\mu = \varphi$ component of the conservation law (\ref{EM-cons})
is written as 
\begin{equation}
 \label{EMconsphi}
  \partial_r \left(\sqrt{-g}T^r_\varphi\right) 
  + \partial_\theta \left(\sqrt{-g} T^\theta_\varphi\right) = 0.
\end{equation}
Noting  $\sqrt{-g} = \left(r^2 + a^2\cos^2\theta\right)\sin\theta$,
we define a spacial component of $U(1)$ current 
$J^r_{(2)}$ for each partial wave mode as follows, 
\begin{equation}
 \label{def-jr}
 J^r_{(2)}(r) \equiv - \int d\Omega_2 \left(r^2 + a^2\cos^2\theta\right) 
  T^r_\varphi.
\end{equation}
Then by integrating the eq.(\ref{EMconsphi}) over the angular coordinates 
the $U(1)$ current is shown to be conserved,
\begin{equation}
\partial_r J^r_{(2)}=0.
\end{equation}

Second, the $\mu=t$ component  $\nabla_\nu T^\nu_t = 0$ becomes  
a modified conservation law of the two-dimensional energy-momentum tensor.
It is written as
\begin{equation}
 \frac{1}{r^2 + a^2\cos^2\theta}
  \partial_r \left[\left(r^2 + a^2\cos^2\theta\right) 
	     \left(T^r_t - A_t T^r_\varphi\right)\right]
  + F_{rt} T^r_\varphi = 0,
\end{equation} 
where $F_{rt} = \partial_r A_t$ 
with the gauge potential $A_t$ defined in eq.(\ref{2d-bg}).  By defining the
energy-momentum tensor $T^r_{t(2)}$ of the two-dimensional effective theory
from $T^r_t$ as 
\footnote{We would like to thank K. Umetsu for pointing out
to us mistakes in eqs. (12) and (13) in the first version of this paper.}
\begin{equation}
 T^r_{t(2)} = \int 
 d\Omega_2 \left(r^2 + a^2 \cos^2\theta\right)
  \left(T^r_t - A_t T^r_\varphi\right),
\end{equation}
we find the following conservation law 
with the $U(1)$ gauge field background $A_t$, 
\begin{equation}
 \partial_r T^r_{t(2)} - F_{rt} J^r_{(2)} = 0.
\end{equation}
This equation is the conservation law for the energy-momentum 
tensor in an electric field background.
(See eq.(18) in our previous paper \cite{Iso:2006wa}).


\section{Anomalies and Hawking Fluxes \label{AHF}}

As explained in the introduction, if we neglect
 classically irrelevant ingoing modes near the
horizon, the effective two-dimensional 
theory becomes chiral near the horizon and 
the gauge symmetry or the general coordinate
covariance becomes anomalous due to
the gauge or gravitational anomalies.

The following procedure to obtain the Hawking
fluxes from the anomalies is parallel to the 
analysis for Reissner-Nordstr\"om
black holes  \cite{Iso:2006wa}. 

First we determine the flux of the $U(1)$ current.
In the $d=4$ language, the $U(1)$ flux corresponds to the
flux of angular momentum carried by Hawking
radiation from rotating black holes.
The effective theory outside the horizon $r_+$ is
defined in the region $r \in [r_+,  \infty]$. 
We will divide the region into two. One is a near
horizon region where we neglect the ingoing modes
since such modes never come out once 
they fall into  black holes. The other region is
apart from the horizon.
The current is conserved 
\begin{equation}
\partial_r J_{(o)}^r =0,
\end{equation} 
in the latter region.
On the contrary, in the near horizon
region $r\in [r_+, r_+ + \epsilon]$, 
since there are only outgoing
(right handed) fields, the current obeys an
 anomalous equation 
\begin{equation}
 \partial_r J^r_{(2)} = \frac{m^2}{4\pi}\partial_r A_t.
\end{equation}
The right hand side is a gauge anomaly in a 
consistent form 
\cite{Bertlmann:xk}\cite{Fujikawa:2004cx}\cite{Bardeen:1984pm}.
The current is accordingly a 
 consistent current which can be obtained from the variation 
of the effective action with respect to the gauge potential.
We can solve these equations in each region as
\begin{eqnarray}
J_{(o)}^r &=& c_o, \\
J_{(H)}^r &=& c_H + \frac{m^2}{4\pi}
\left( A_t(r) -A_t(r_+) 
\right),
\end{eqnarray}
where $c_o$ and $c_H$ are integration constants.
$c_o$ is the value of the current at $r=\infty$. 
$c_H$ is the value of the consistent current of the outgoing modes
at the horizon.  
Current is written as a sum in two regions
\begin{equation}
J^{\mu} = J_{(o)}^{\mu} \Theta_+(r) +  J_{(H)}^{\mu} H(r),
\label{outgoingcurrent}
\end{equation}
where $\Theta_+(r) = \Theta(r-r_+-\epsilon)$ and $H(r)=1-\Theta_+(r)$
are step functions defined in the region $r \in [r_+, \infty ].$
Note that since we have neglected the ingoing modes near the horizon
this current is only a part of the total current. 
The total current including a contribution from the near horizon ingoing
modes is given by
\begin{equation}
J^{\mu}_{total} = J^{\mu} + K^\mu
\end{equation}
where 
\begin{equation}
K^\mu = -\frac{m^2}{4 \pi} A_t(r) H(r).
\label{ingoingcurrent}
\end{equation}
This cancels the anomalous part in $J^{\mu}$ near the horizon.

We now consider the effective action $W$ where we have neglected the
classically irrelevant ingoing modes at the horizon. Hence the variation of
the effective action under gauge transformations is given by  
\begin{equation}
-\delta W = \int d^2 x \sqrt{-g_{(2)}} \lambda 
\nabla_{\mu} J^{\mu}
\end{equation}
where $\lambda$ is a gauge parameter.
By integration by parts we have
\begin{equation}
 - \delta W = \int d^2 x \lambda 
  \left[  
   \delta(r-r_+ - \epsilon) 
   \left(J_{o}^r - J_H^r + \frac{m^2}{4 \pi}A_t \right) 
   + \partial_r \left(\frac{m^2}{4\pi}A_t H\right)
  \right].
\end{equation}
As well as the current (\ref{outgoingcurrent}), this effective action 
does not contain a contribution from the near horizon ingoing modes. 
The total effective action must be gauge invariant and the last term should
be cancelled by quantum effects of the classically irrelevant ingoing
modes. Namely a contribution from
the ingoing modes  (\ref{ingoingcurrent}) cancels
the last term. 
The coefficient of the delta-function should also vanish, which relates the
coefficient of the current in two regions; 
\begin{equation}
c_o = c_H - \frac{m^2}{4\pi} A_t(r_+).
\end{equation}
This relation ensures that the total current 
$J^\mu_{total}$ is conserved in all the regions;
  $\partial_r J^r_{total}=0.$

In order to fix the value of the current, we 
impose that the coefficient of the covariant current at the
horizon should vanish. 
This assumption is based on the following physical
requirement. In the near horizon region, we have first neglected
ingoing modes. Hence the current there has contributions from only
the outgoing modes which depend on $u=t-r_*$. 
Namely the vanishing condition for the covariant current is nothing
but the vanishing condition for the current of the outgoing modes, 
which is usually imposed to assure regularity of 
the physical quantities at the future horizon.
We will discuss it more in the discussions and in Appendix \ref{A2}. 
Another condition we have implicitly assumed is the constant value of the
ingoing current (\ref{ingoingcurrent}). We could have added an arbitrary
constant in (\ref{ingoingcurrent}).  
The boundary condition for $K^\mu$ to vanish
at $r=\infty$ corresponds to a condition that there is no ingoing modes
at radial infinity.

Since the covariant current $\tilde{J^r}$ 
is written as 
 $\tilde{J^r} = J^r + \frac{m^2}{4\pi} A_t(r) H(r)$, 
the condition $\tilde{J}^r(r_+)=0$  determines the value of the charge flux
to be 
\begin{equation}
 c_o = - \frac{m^2}{2 \pi} A_t(r_+) = \frac{m^2 a}
  {2 \pi (r_+^2 +a^2)}.
\label{Jflux}
\end{equation}
This agrees with the flow of the angular momentum 
associated with the Hawking thermal (blackbody) radiation.
(See  eq. (\ref{BBangular}) in the appendix, with $Q=0$.)

Similarly we can determine the flux of the 
energy-momentum tensor radiated from 
Kerr black holes.
Since there is an effective background gauge potential, 
the energy-momentum tensor satisfies the 
modified conservation equation outside the horizon:
\begin{equation}
 \partial_r T^r_{t (o)} =  F_{r t }  J^r_{(o)}.
\end{equation}
By using $J_{(o)}^r=c_o$ it is solved as
\begin{equation}
 T^r_{t(o)}=a_o + c_o  A_t(r)
\end{equation}
where $a_o$ is an integration constant. This is the value of the 
energy flow at $r=\infty$. 
In the near horizon region, there are gauge and 
gravitational anomalies and the conservation
equation is modified as
\begin{equation}
 \partial_r T^r_t = F_{rt}J^r + A_t \nabla_\mu J^\mu + \partial_r N^r_t,
\end{equation}
where  $ N^r_t =( f^{\prime 2}+f f^{\prime\prime})/192\pi.$ 
(Refer to \cite{Iso:2006wa} for the derivation.)
The second term
comes from the gauge 
anomaly while the third one is the gravitational
anomaly for the consistent energy-momentum tensor \cite{Alvarez-Gaume:1983ig}.
The first and the second term can be combined in terms of the covariant
current $\tilde{J}_{(H)}^r$ as $F_{rt} \tilde{J}_{(H)}^r $.
By substituting 
$\tilde{J}_{(H)}^r = c_o + \frac{m^2}{2\pi} A_t(r)$ into this equation, 
$T^r_{t{(H)}}$ can be solved as
\begin{equation}
 T^r_{t{(H)}} = a_H + 
 \int^r_{r_+} dr \partial_r \left(
 c_o A_t + \frac{m^2}{4\pi}A_t^2 + N^r_t
 \right).
 \end{equation}
The energy-momentum tensor  combines contributions from these two regions, 
$T^\mu_{\nu} = T^\mu_{\nu~(o)}\Theta_+ + T^\mu_{\nu~(H)} H$. 
This does not 
contain a contribution from the ingoing modes near the horizon.
The total energy-momentum tensor is a sum of $T^\mu_\nu$ and 
$U^\mu_\nu$,  where 
\begin{equation}
U^r_t =-\left(\frac{m^2}{4\pi}A_t^2(r) + N^r_t(r)\right)H,
 \label{ingoingEM}
\end{equation}
is a contribution from the ingoing modes. 
The freedom to add a constant value is fixed by a requirement
that it should vanish at $r=\infty.$ This condition corresponds to 
a condition that there is no ingoing energy flow at 
$r=\infty.$

Under the following diffeomorphism transformation with a transformation
parameter $\xi^t$, the effective action
(without the near horizon ingoing modes) changes as
\begin{eqnarray}
 - \delta W &=& \int d^2x \sqrt{-g_{(2)}} ~\xi^t \nabla_\mu T^\mu_{t} 
  \nonumber \\
 &=& \int d^2x ~\xi^t
  \left[c_o \partial_r A_t(r) +
   \partial_r \left( \frac{m^2}{4\pi} A_t^2 + N^r_t \right)
  \right. \nonumber \\
 && \left. \hspace{20mm}
      +\left(T^r_{t~(o)} - T^r_{t~(H)} 
       + \frac{m^2}{4\pi}A_t^2+N^r_t\right) \delta(r-r_+ -\epsilon) 
     \right].
\end{eqnarray}
The first term is the classical effect of the background electric field for
constant current flow. The second term should be cancelled by the quantum
effect of the ingoing modes (\ref{ingoingEM}). The coefficient of the last
term should vanish 
in order to restore the diffeomorphism covariance at the horizon.  This
relates the coefficients: 
\begin{equation}
 a_o = a_H +\frac{m^2}{4\pi}A_t^2(r_+)- N^r_t(r_+) . 
 \end{equation}

In order to determine $a_o$,  we impose a vanishing
condition for the covariant energy-momentum tensor at the horizon. 
This condition corresponds to the regularity condition for the 
energy-momentum tensor at the future horizon.
Since
the covariant energy-momentum tensor is related to the consistent one by 
\cite{Bardeen:1984pm}\cite{Bertlmann:2000da}
\begin{equation}
\tilde{T}^r_t = T^r_t +\frac{1}{192\pi} (f f'' -2(f')^2),
\end{equation} 
the condition reads 
\begin{equation}
a_H= \kappa^2/24 \pi = 2N^r_t(r_+),
\end{equation}  
where 
\begin{equation}
\kappa=2 \pi /\beta =\frac{1}{2}\partial_r 
f \Big|_{r=r_+}
=\frac{r_+ - r_-}{2(r_+^2 + a^2)}
\end{equation}
 is the
surface gravity of the black hole.
The total flux of the energy-momentum tensor is given by
\begin{equation}
a_o=\frac{m^2 a^2}{4\pi (r_+^2+a^2)^2} +N^r_t(r_+)
=\frac{m^2 \Omega^2}{4\pi } +\frac{\pi }{12 \beta^2},
\label{EMflux}
\end{equation}
where $\Omega$ is an angular velocity at the horizon,
\begin{equation}
 \label{def-Omega}
 \Omega = \frac{a}{r_+^2 + a^2}.
\end{equation}
This value of the flux is the same as the Hawking
flux from Kerr black holes in eq.(\ref{BBenergy}) with $Q=0.$


\section{Quantum fields in Kerr-Newman black hole \label{KN}}
In this section we generalize our analysis to rotating
charged (Kerr-Newman) black holes and obtain 
Hawking fluxes. 
The analysis in the previous section can be straightforwardly applied to
this case. 

The metric of the Kerr-Newman black hole is given by replacing
$\Delta$ in (\ref{Kerrmetric}) with 
\begin{equation}
 \Delta = r^2 -2Mr +a^2 + Q^2 = (r-r_+)(r-r_-),
\end{equation}
where $Q$ is the electric charge of the black hole and $r_{+(-)}$ are radii
of outer (inner) horizons
\begin{equation}
 r_\pm = M \pm \sqrt{M^2 -a^2 -Q^2}.
\end{equation}
The background gauge field is given by
\begin{equation}
 A = -\frac{Qr}{r^2 + a^2\cos^2\theta}
  \left(dt - a\sin^2\theta d\varphi\right).
\end{equation}
Let us consider a complex scalar field in this background.
As well as the case of the Kerr black hole background, each partial wave
mode of fields 
can be described near the outer horizon by the following effective 
$(1+1)$-dimensional theory in the $(r-t)$ section,
\begin{eqnarray}
 S = -\int dt dr ~(r^2 + a^2)\phi^*_{lm} 
  \left[
   \frac{r^2 + a^2}{\Delta}
   \left(\partial_t + \frac{ieQr}{r^2+a^2} + \frac{iam}{r^2+a^2}\right)^2 
   -\partial_r \frac{\Delta}{r^2 + a^2} \partial_r
  \right] \phi_{lm},
\end{eqnarray} 
where $e$ is the electric charge of $\phi$.
The dilaton background and the metric have the same forms
as the ones of the Kerr geometry (\ref{2d-bg}).
 $U(1)$ gauge field background is now given by 
\begin{eqnarray}
 {\cal A}_t = - \frac{eQr}{r^2+a^2} - \frac{ma}{r^2 + a^2}.
\label{totalpotential}
\end{eqnarray}
The first term is originated from the electric field of the Kerr-Newman
black hole while the second one is the induced gauge potential from the
metric which is  associated with the axisymmetry of
the Kerr-Newman background. 

In this case, there are two $U(1)$ gauge symmetries
and correspondingly two gauge currents. 
One is the original gauge symmetry while
the other is the induced gauge symmetry 
associated with the isometry along the $\varphi$
direction. 
The gauge potential
(\ref{totalpotential}) is a sum of these two fields,
\begin{equation}
{\cal A}_t = e A_t^{(1)}+m A_t^{(2)}.
\label{potentialsum}
\end{equation}
 The $U(1)$ current $j^r$ associated with the original gauge symmetry
 is defined from the
electric current $J^r$ in the four-dimensional spacetime as 
\begin{eqnarray}
 j^r = \int d\Omega_2 (r^2 + a^2\cos^2\theta) J^r.
\end{eqnarray}
Since the background is time-independent, the current in the Kerr-Newman
background satisfies 
\begin{eqnarray}
 \partial_r j^r = 0.
\end{eqnarray}
In the region $r \in [r_+, r_+ + \epsilon]$, this equation is modified by
the gauge anomaly,
\begin{eqnarray}
 \partial_r j^r = \frac{e}{4\pi}\partial_r {\cal A}_t.
\end{eqnarray}
Following the procedure in the case of Kerr black hole, 
we can obtain the flux of the electric charge as
\begin{equation}
 -\frac{e}{2\pi} {\cal A}_t(r_+) 
  = \frac{e}{2\pi}
  \left(\frac{eQr_+}{r_+^2+a^2} + \frac{ma}{r_+^2 + a^2}\right).
\end{equation}
This reproduces the flux of the electric current derived from the Hawking
radiation in (\ref{BBcharge}). 

Next the current $J^r_{(2)}$ associated with the axial symmetry can be
defined from  the $(r, \phi)$-component of the four-dimensional 
energy-momentum tensor $T^r_\varphi$ as eq.(\ref{def-jr}).
The anomalous equation near the horizon is
\begin{equation}
 \partial_r J^r_{(2)} = \frac{m}{4\pi}\partial_r {\cal A}_t.
\end{equation}
Hence the flux of the angular momentum is obtained as
\begin{equation}
 -\frac{m}{2\pi} {\cal A}_t(r_+) 
  = \frac{m}{2\pi}
  \left(\frac{eQr_+}{r_+^2+a^2} + \frac{ma}{r_+^2 + a^2}\right),
\end{equation}
which is equal to (\ref{BBangular}). 
It should be noted that $j^r$ and $J^r_{(2)}$ 
are not independent for a fixed azimuthal angular momentum $m$.
Actually as is clear from the gauge potential (\ref{potentialsum})
their expectation value in the Kerr-Newman background are related as
$\frac{1}{e}j^r = \frac{1}{m} J^r_{(2)} ( \equiv {\cal J}^r)$.

Finally the anomalous equation for the energy-momentum tensor in the region 
$r \in [r_+, r_+ + \epsilon]$ is given by
\begin{equation}
 \partial_r T^r_t = {\cal F}_{rt}{\cal J}^r 
  + {\cal A}_t\partial_r {\cal J}^r
  + \partial_r N^r_t,
\end{equation}
where ${\cal F}_{rt} = \partial_r {\cal A}_t$.
${\cal J}^\mu$ is defined above and satisfies
$\partial_r {\cal J}^r = \frac{1}{4\pi}\partial_r {\cal A}_t$.
Applying the same method as in the previous section,
the flux of the energy-momentum is determined as
\begin{equation}
 \frac{1}{4\pi} {\cal A}_t^2(r_+) + N^r_t(r_+)
  = \frac{1}{4\pi}
  \left(\frac{eQr_+}{r_+^2+a^2} + \frac{ma}{r_+^2 + a^2}\right)^2
  + \frac{\pi}{12\beta^2},
\end{equation}
where $\beta$ is the Hawking temperature of the Kerr-Newman black hole,
\begin{equation}
 \frac{2\pi}{\beta} = \frac{r_+ - r_-}{2(r_+^2 + a^2)}.
\label{beta}
\end{equation}
This is the flux of energy expected from the Hawking radiation
(\ref{BBenergy}). 

\section{Conclusions and Discussions \label{CD}}
In this paper, we extended our previous analysis of Hawking radiation from
charged black holes based on gauge and gravitational anomalies
to  the cases of rotating black holes, i.e. Kerr and Kerr-Newman black
holes. 
In the case of Hawking radiations from Kerr black hole, though there is no
gauge symmetry in the original four-dimensional setting, 
the technique for Reissner-Nordstr\"om black hole can be utilized since the
effective two-dimensional theory near the horizon can be described by 
charged matter fields in an electric field. This is because the axial
direction of the four-dimensional general coordinate transformations can be
interpreted as $U(1)$ gauge symmetry for each partial mode. 
The charge of the field is given by the azimuthal quantum number.
By this identification, we have reproduced the correct Hawking flux from
Kerr black holes by demanding gauge and diffeomorphism symmetry. This
analysis was straightforwardly extended to the Hawking radiations of charged
particles from Kerr-Newman black hole. 

The derivation is based only on the anomaly equation for gauge current and
energy-momentum tensor in effective two-dimensional field theories near
horizons and the result is universal. 
Namely it does not either depend on the detailed dynamics of fields apart
from the horizon or the spin of the radiated particles. Of course, when
these radiated particles travel to the infinity, they experience potentials
or interactions and the spectrum is modified. Our treatment considered only
the near horizon effect and neglected such the scattering effect outside the
horizon (i.e. grey body factor). 

Our derivation is partial since we have not been able to derive the
frequency-dependent spectrum of the Hawking radiation. For this purpose, we
may need to develop frequency dependent formulation of anomalies or
renormalization group type analysis near the horizon. 
This is left for future investigation.

Finally we would like to comment on our choice of boundary conditions
and the regularity of the physical quantities 
at the future horizon. As is well known, Hawking radiation is derived 
by assuming regularity of the energy-momentum tensor at the future horizon
and an assumption that there is no ingoing current at the past horizon 
\cite{Unruh:db}.
This boundary condition corresponds to the Unruh vacuum, and
fluxes for other vacua correspond to other choices of
boundary conditions. In the case of Reissner-Nordstr\"om black hole
or rotating black holes, each partial mode of 
four-dimensional fields is effectively
described by a massless free two-dimensional conformal field
in an electric and gravitational background.
Hence we can calculate the effective action exactly and the
currents or energy-momentum tensor are also exactly obtained
up to boundary conditions of Green functions. (see Appendix \ref{A2}.) 
If we impose a regularity
at the future horizon and absence of ingoing fluxes at $r=\infty$, we
can obtain the fluxes for Unruh vacuum. 
(In the Schwarzschild case, see a review \cite{wipf}.)
We have chosen the 
boundary condition that the radial component of the covariant current
should vanish at the horizon. 
This corresponds to the above regularity condition in the following
sense. Near the horizon, we have first neglected the quantum effect
of ingoing modes. Hence the current in the near horizon region 
should be considered as the outgoing current. In the $(u,v)$ coordinates
where $u=t-r_*$ and $v=t+r_*$, the vanishing condition for the covariant
current corresponds to the condition $J_u \rightarrow 0$ at the future
horizon. 
This is the regularity condition at the future horizon. 
On the other hand, the boundary condition for ingoing modes at infinity 
is implicitly assumed. 
We have derived the Hawking flux by using anomalies and conservation laws
for currents or energy-momentum tensor. But there is  a freedom to add an
extra constant ingoing flux in the whole region because such an addition
does not break the conservation laws. 
(Such a constant cannot be added to the outgoing flux 
 because this addition violates the
regularity at the future horizon.)  
We have taken  into account the quantum effect of ingoing modes through
the anomalous contribution (WZ term). This corresponds to the boundary
condition that the ingoing modes should vanish at infinity.

\begin{acknowledgments}
We would like to thank Drs. K. Fujikawa, H. Kawai, H. Kodama, M. Natsuume, 
T. Yoneya and other participants of KEK theory
workshop (March 13 - 16, 2006) for discussions and comments.
 
After completing the paper, we are informed of a similar work 
independently done by  K. Murata and J. Soda \cite{soda}.   

This work is supported in part by funds provided by the U.S. Department
of Energy (D.O.E.) under cooperative research agreement
DE-FC02-94ER40818. 
\end{acknowledgments}

\appendix

\section{Blackbody radiation \label{A1}}
In the body of the paper, we have treated a scalar
field in rotating black holes. The same trick to reduce the
system to effective $d=2$ theory  
can be applied to  fermions with a slightly more complication.
In this appendix we calculate the flux of Hawking 
radiations in the case of fermions in order to avoid
the problem of superradiance. 
The  Hawking distribution is given by the Planck distribution with 
chemical potentials for an azimuthal angular momentum $m$ and an electric
charge $e$ of the fields radiated from the black hole.
For fermions the distribution for the Kerr-Newman black hole is given by
\begin{eqnarray}
  N_{e, m}(\omega) 
   = \frac{1}{e^{\beta(\omega - e\Phi - m\Omega)}+1},
\end{eqnarray}
where $\Phi=Qr_+/(r_+^2+a^2)$ and $\Omega$ 
was defined in eq.(\ref{def-Omega}). The inverse 
temperature $\beta$ is defined in (\ref{beta}).
From this distribution, we can calculate 
fluxes of the electric current $j$, 
angular momentum $J_{(2)}$ and energy-momentum tensor,
defined respectively as $F_Q$, $F_a$ and $F_M$;
\begin{eqnarray}
 F_Q &=& 
  e \int_0^\infty \frac{d\omega}{2\pi}
  \left(N_{e,m}(\omega) - N_{-e, -m}(\omega)\right)
  = \frac{e}{2\pi}\left(e\Phi + m\Omega\right), 
  \label{BBcharge} \\
 F_a &=& 
  m \int_0^\infty \frac{d\omega}{2\pi}
  \left(N_{e,m}(\omega) - N_{-e, -m}(\omega)\right)
  = \frac{m}{2\pi}\left(e\Phi + m\Omega\right), 
  \label{BBangular} \\
 F_M &=& 
  \int_0^\infty \frac{d\omega}{2\pi}~\omega
  \left(N_{e,m}(\omega) + N_{-e, -m}(\omega)\right)
  = \frac{1}{4\pi}\left(e\Phi + m\Omega\right)^2+ \frac{\pi}{12 \beta^2}.
  \label{BBenergy}
\end{eqnarray}
Here we added contributions from a particle with a quantum number $(e, m)$ 
and its anti-particle with $(-e, -m)$ in order to compare with our results.


\section{Effective action and Hawking radiation \label{A2}}

It is shown that  each partial wave mode in black hole backgrounds can be
described near the outer horizon by a two-dimensional effective theory.
Since mass, potential and interaction terms can be neglected near the
horizon, the effective $d=2$ theories are free conformal theories.
Thus we can evaluate fluxes of the current or energy by calculating 
the effective action directly.
Here we will calculate such fluxes in the Reissner-Nordstr\"om case
because Kerr or Kerr-Newman cases are reduced to the same
calculation as explained in the body of the paper.

In the Schwarzschild case, many works have been done to derive 
Hawking flux from effective actions in black hole background
\cite{wipf, regularity}.  Since the effective $d=2$ theories contain
dilaton background, it seems  necessary to include the effect
of dilaton in such investigations and there have been many 
discussions on it.
However, as we have briefly commented
in our previous paper \cite{Iso:2006wa}, the effect of dilaton does not
change the property of $(r,t)$-component of the energy-momentum tensor
and accordingly the Hawking flux is independent of the dilaton background. 
It only affects the other nonuniversal components like $T^t_t$.
In this sense, Hawking radiation is universal. Once we impose
the boundary conditions, the value of the flux is determined only by the 
value of anomalies at the horizon. 
Therefore, in the following, we calculate  fluxes of current and energy in
the Reissner-Nordstr\"om black hole for free massless scalar fields without
dilaton backgrounds.

The gauge potential and the metric of the Reissner-Nordstr\"om black hole is 
given by
\begin{eqnarray}
 A &=& -\frac{Q}{r}dt, 
 \label{potentialA} \\
 ds^2 &=& f(r)dt^2 - \frac{dr^2}{f(r)}+r^2 d\Omega^2,
\end{eqnarray}
where $f(r)$ is
\begin{eqnarray}
 f = 1-\frac{2M}{r}+\frac{Q^2}{r^2} = \frac{(r-r_+)(r-r_-)}{r^2}.
\end{eqnarray}
$r_\pm = M \pm \sqrt{M^2-Q^2}$ are the radii of the outer and inner
horizons.
We also use other coordinate system $(u, v)$,
\begin{eqnarray}
 v=t+r_*, \qquad u=t-r_*, \qquad (dr_* = \frac{1}{f}dr)
\end{eqnarray}
and the Kruskal coordinates,
\begin{eqnarray}
 U=-e^{-\kappa_+ u}, \qquad V=e^{\kappa_+ v},
\end{eqnarray}
where $\kappa_+$ is the surface gravity on the outer horizon,
$\kappa_+=\frac{r_+ - r_-}{2r_+^2}$.

Each partial wave of charged matter fields in the Reissner-Nordstr\"om black
hole background is effectively described by a charged field in a $d=2$
charged black hole with a metric 
\begin{equation}
ds^2 =f(r)dt^2 - \frac{dr^2}{f(r)}
\end{equation}
and the gauge potential (\ref{potentialA}). 
The two-dimensional curvature  obtained from this metric is given by
\begin{equation}
R_{tt}=\frac{f f''}{2}, \ \ R_{rr}=-\frac{f''}{2f'}, 
 \ \ R_{tr}=0, \ \ R=f''. 
 \label{R}
\end{equation}

Effective action $\Gamma$ of a conformal field with a central charge $c=1$
in this gravitational and electric field background consists of the
following two parts; the gravitational part $\Gamma_{grav}$ and 
gauge field part $\Gamma_{U(1)}$.
The gravitational part (Polyakov action) is given by,
\begin{equation}
 \Gamma_{grav} = \frac{1}{96\pi}\int d^2x d^2y
  \sqrt{-g}R(x)\frac{1}{\triangle_g}(x, y) \sqrt{-g}R(y),
\end{equation}
while the $U(1)$ gauge field part is
\begin{equation}
 \Gamma_{U(1)} = \frac{e^2}{2\pi}\int d^2x d^2y
  ~\epsilon^{\mu\nu}\partial_\mu A_\nu(x) 
  \frac{1}{\triangle_g}(x, y)\epsilon^{\rho\sigma}\partial_\rho
  A_\sigma(y). 
\end{equation}
$R$ is the two-dimensional scalar curvature (\ref{R}) and $\triangle_g$ is 
the Laplacian in this background.
From these effective actions, we can obtain the energy-momentum tensor
$T_{\mu\nu}$ and $U(1)$ current $J^\mu$ 
(see \cite{Leutweyler} for a chiral case),
\begin{eqnarray}
 T_{\mu\nu} &=& T_{\mu\nu}^{grav} + T_{\mu\nu}^{U(1)} 
 = \frac{2}{\sqrt{-g}}\frac{\delta \Gamma}{\delta g^{\mu\nu}}, \\
 T_{\mu\nu}^{grav} &=& \frac{1}{48\pi}
  \left(
   2g_{\mu\nu}R -2\nabla_\mu\nabla_\nu S + \nabla_\mu S \nabla_\nu S
   -\frac{1}{2}g_{\mu\nu}\nabla^\rho S \nabla_\rho S
  \right), \\
 T_{\mu\nu}^{U(1)} &=& \frac{e^2}{\pi}
  \left(
   \nabla_\mu B \nabla_\nu B 
   -\frac{1}{2} g_{\mu\nu}\nabla^\rho B \nabla_\rho B
  \right), \\
 \label{u(1)current}
 J^\mu &=& \frac{1}{\sqrt{-g}}\frac{\delta\Gamma}{\delta A_\mu} 
  = \frac{e^2}{\pi}\frac{1}{\sqrt{-g}} \epsilon^{\mu\nu}\partial_\nu B,
\end{eqnarray}
where
\begin{eqnarray}
 S(x) &=& \int d^2y \frac{1}{\triangle_g}(x, y)\sqrt{-g}R(y), \\
 B(x) &=& \int d^2y \frac{1}{\triangle_g}(x, y)
  \epsilon^{\mu\nu}\partial_\mu A_\nu(y).
\end{eqnarray}
Hence $B$ is a solution of the equation,
\begin{equation}
 \triangle_g B = \epsilon^{\mu\nu}\partial_\mu A_\nu
  = -\partial_r A_t (r).
\end{equation}
This equation is solved as
\begin{eqnarray}
 B &=& B_0 + b(u) + \tilde{b}(v), \\
 && \partial_r B_0 = \frac{1}{f} \left(A_t (r) + c\right),
\end{eqnarray}
where $b(u)$ and $\tilde{b}(v)$ are solutions of the homogeneous equation 
$\triangle_g B = \frac{4}{f}\partial_u \partial_v B =0$, and $c$ is an
integration constant. 
Thus the electromagnetic current (\ref{u(1)current}) becomes
\begin{eqnarray}
 J^t &=& \frac{e^2}{\pi}\partial_r B 
  = \frac{e^2}{\pi f}\left(A_t (r) + c\right)
  + \frac{e^2}{\pi}\partial_r (b(u) + \tilde{b}(v)),\\
 J^r &=& - \frac{e^2}{\pi}\partial_t B
  = -\frac{e^2}{\pi}\partial_t (b(u) + \tilde{b}(v)).
\end{eqnarray}
In the $(u, v)$ coordinates they are given by
\begin{eqnarray}
 J_u &=& \frac{f}{2}\left(J^t - \frac{1}{f}J^r\right)
  = \frac{e^2}{2\pi}\left(A_t (r) + c\right)
  - \frac{e^2}{\pi}\partial_u b(u),\\
 J_v &=& \frac{f}{2}\left(J^t + \frac{1}{f}J^r\right)
  = \frac{e^2}{2\pi}\left(A_t (r) + c\right) 
  + \frac{e^2}{\pi}\partial_v \tilde{b}(v).
\end{eqnarray}
In order to determine the homogeneous parts, we impose
the following boundary conditions.
First we require that free falling observers see a finite 
(not infinite) amount of the charged
current at the outer horizon and accordingly the current
in the Kruskal coordinate $U$ is required to be finite at the 
future horizon.  
Since $J_U = -(1/\kappa_+ U) J_u$ and 
$U \rightarrow \sqrt{r-r_+} \ (r \rightarrow r_+)$, 
$J_u$ must vanish on the horizon, 
\begin{eqnarray}
 J_u \ \mathop{\longrightarrow}^{r\rightarrow r_+} \ 
  \frac{e^2}{2\pi}\left(A_t (r_+) + c\right) 
  -\frac{e^2}{\pi}\partial_u b(u)\Big|_{r=r_+} = 0.
\end{eqnarray}
This determines the homogeneous part $b(u)$ as 
$\partial_u b(u) = \frac{1}{2}\left(A_t (r_+) + c\right)$.
Second we impose that there is no ingoing current at $r=\infty$ and require 
\begin{eqnarray}
 J_v \ \mathop{\longrightarrow}^{r\rightarrow \infty} \ 
  \frac{e^2}{2\pi}c 
  + \frac{e^2}{\pi}\partial_v \tilde{b}(v)\Big|_{r\rightarrow\infty} = 0.
\end{eqnarray}
This determines the other homogeneous part $\tilde{b}(v)$ as
$\partial_v \tilde{b}(v) = -c/2$.
By these boundary conditions the $U(1)$ current is completely determined as
\begin{eqnarray}
 J_u &=& \frac{e^2}{2\pi}\left(A_t (r) - A_t (r_+)\right),\\
 J_v &=& \frac{e^2}{2\pi} A_t (r).
\end{eqnarray}
In $(t,r)$-coordinate, the $U(1)$ current is given by
\begin{eqnarray}
 J^r &=& J_u - J_v = -\frac{e^2}{2\pi}A_t (r_+)
  = \frac{e^2Q}{2\pi r_+}, \\
 J^t &=& \frac{1}{f}\left(J_u + J_v\right)
  = \frac{e^2}{\pi f}\left(A_t(r) - \frac{1}{2}A_t(r_+)\right).
\end{eqnarray}
This is the expectation value of the current for the Unruh vacuum 
in the $d=2$ Reissner-Nordstr\"om black hole.

The energy-momentum tensor can be similarly obtained.
$S(x)$ satisfies the equation
\begin{equation}
\triangle_g S =R=f''
\end{equation}
and it can be solved to be a sum of an inhomogeneous  and 
homogeneous parts.

Hence the energy-momentum tensor is written as
\begin{eqnarray}
 T_{uu} &=& \frac{1}{192\pi}\left(-f'^2 + 2 f f''\right)
  +\frac{e^2}{4\pi}\left(A_t (r) - A_t (r_+)\right)^2 
  + t(u), \\
  T_{vv} &=& \frac{1}{192\pi}\left(-f'^2 + 2 f f''\right)
  +\frac{e^2}{4\pi}A_t^2 (r)
  + \tilde{t}(v), \\
 T_{uv} &=& \frac{1}{96\pi}ff'',
\end{eqnarray}
where $t(u) ~(\tilde{t}(v))$ is an arbitrary function of $u ~(v)$ that
are determined by boundary conditions.
Similarly to the $U(1)$ current 
we impose the following boundary conditions,
\begin{eqnarray}
 && T_{uu} \ \mathop{\longrightarrow}^{r\rightarrow r_+} \ 
  -\frac{1}{192\pi}f'^2(r_+) + t(u)\Big|_{r=r_+} = 0, \\
 && T_{vv} \ \mathop{\longrightarrow}^{r\rightarrow\infty} \ 
  \tilde{t}(v)\Big|_{r\rightarrow\infty} = 0.
\end{eqnarray}
Then components of the energy-momentum tensor become
\begin{eqnarray}
 T^t_t &=& \frac{1}{f}\left(T_{uu} + T_{vv} + 2T_{uv}\right) \nonumber \\
 &=& \frac{1}{96\pi f}
  \left[-(f')^2 + 4ff'' + \frac{1}{2}(f'(r_+))^2
  \right]
  +\frac{e^2}{2\pi f}
  \left[A_t^2 - A_t(r_+)A_t + \frac{1}{2}A_t^2(r_+)\right], \\
 T^r_r &=& -\frac{1}{f}\left(T_{uu} + T_{vv} - 2T_{uv}\right) \nonumber \\
 &=& \frac{f}{96\pi}\left[(f')^2 -\frac{1}{2} f'(r_+)^2\right]
  - \frac{e^2}{2\pi f}
  \left[A_t^2 - A_t(r_+)A_t + \frac{1}{2}A_t^2(r_+)\right], \\
 T^r_t &=& T_{uu} - T_{vv} 
  = -\frac{e^2}{2\pi}A_t(r_+)A_t(r) + \frac{1}{192\pi}f'^2(r_+)
  + \frac{e^2}{4\pi}A_t^2(r_+).
\end{eqnarray}
Therefore the flux of the energy $T^r_t$ is obtained as
\begin{eqnarray}
 T^r_t  \ \mathop{\longrightarrow}^{r\rightarrow \infty}  \ 
  \frac{1}{192\pi}f'^2(r_+) + \frac{e^2}{4\pi}A_t^2(r_+).
\end{eqnarray}

\section{Notations for the Reduction  \label{A3}}
In this appendix we summarize our notations for the Kaluza-Klein 
reductions of the metrics and conservation equations
\footnote{This appendix was added after the paper has been published in Physical Review D
in order to clarify some confusions in the conservation equations in $d=2$.}.
\par
We first divide the $d$-dimensional coordinates into $X^A = \{ x^\mu, \phi^i\}$
and write the original metric as  $\hat{g}_{AB}$.
The metric is written as usual in the form of the  Kaluza-Klein:
\begin{eqnarray}
(\hat{g}_{AB}) &=& 
  \left(
   \begin{array}{cc}
    g_{\mu\nu} + h_{ij} A^{(i)}_\mu A^{(j)}_\nu & h_{jk} A^{(k)}_\mu \\
    h_{ik} A^{(k)}_\nu & h_{ij}
   \end{array}
  \right), \\
  (\hat{g}^{AB}) &=&
  \left(
   \begin{array}{cc}
    g^{\mu\nu} & - A^{(j)\mu} \\
    - A^{(i)\nu} & h^{ij} + g^{\mu\nu} A^{(i)}_\mu A^{(j)}_\nu
   \end{array}
  \right),
\end{eqnarray}
where we  assume that 
$g_{\mu\nu}, A^{(i)}_\mu$ and $h_{ij}$ depend only on $x^\mu$. 
They satisfy
\begin{eqnarray}
 && \hat{g}^{AC} \hat{g}_{CB} = \delta^A_B, \qquad
  g^{\mu\rho} g_{\rho\nu} = \delta^\mu_\nu, \qquad
  h^{ik} h_{kj} = \delta^i_j,  \qquad
  A^{(i)\mu} = g^{\mu\nu} A^{(i)}_\nu.
\end{eqnarray}
\par
The $d$-dimensional coordinate transformations generated
by $\hat{\xi}^A$ can be 
interpreted as the gauge transformation and 
the coordinate transformations on the reduced 
space-time. In the following we assume that the transformation 
parameters are functions of  $x^\mu.$
First if we consider the transformation 
with $\hat{\xi}^A = (0, \xi^i (x)) $, it generates 
gauge transformation as
  \begin{eqnarray}
	\delta g_{\mu\nu} = 0, \qquad
	 \delta A^{(i)}_\mu = \partial_\mu \xi^i, \qquad
	 \delta h_{ij} = 0.
       \end{eqnarray}
Next for the parameter $\hat{\xi}^A = (\xi^\mu (x), 0)$, 
we get the coordinate transformation on the reduced space-time:
 \begin{eqnarray}
	\delta g_{\mu\nu} = \nabla_\mu \xi_\nu + \nabla_\nu \xi_\mu, \qquad
	 \delta A^{(i)}_\mu = \xi^\nu \partial_\nu A^{(i)}_\mu
	 + \partial_\mu \xi^\nu A^{(i)}_\nu, \qquad
	 \delta h_{ij} = \xi^\mu \partial_\mu h_{ij},
       \end{eqnarray}
       where $\xi_\mu = g_{\mu\nu} \xi^\nu$.
\par
From these general formulae we can obtain the conservation 
equations for the reduced energy-momentum tensor and the gauge
current associated with the above gauge transformations.
The definition of the $d$-dimensional energy-momentum tensor
is given by 
\begin{eqnarray}
 \hat{T}_{AB} 
  &\equiv& \frac{2}{\sqrt{-\hat{g}}} \frac{\delta S}{\delta \hat{g}^{AB}},
\end{eqnarray}
where
$ \hat{g} = det (\hat{g}_{AB}) = det (g_{\mu\nu}) det (h_{ij}) = gh.$
When there are no background fields that transform under
the coordinate transformations, it is conserved
\begin{eqnarray}
 \hat{\nabla}_A \hat{T}^A_B &=& \partial_A \hat{T}^A_B
  + \hat{\Gamma}^A_{AC} \hat{T}^C_B - \hat{\Gamma}^C_{AB} \hat{T}^A_C = 0,
\end{eqnarray}
where 
$ \hat{\Gamma}^C_{AB} = \frac{1}{2} \hat{g}^{CD}
  \left(\partial_A \hat{g}_{BD} + \partial_B \hat{g}_{AD}
  - \partial_D \hat{g}_{AB}\right).$
Each component of the conservation equations can be
reduced to the following sets of conservation equations on the
reduced space:
\begin{enumerate}
 \item $\hat{\nabla}_A \hat{T}^A_i = 0$ 
       \begin{eqnarray}
	\frac{1}{\sqrt{-\hat{g}}}
	 \partial_\mu \left(\sqrt{-\hat{g}} \hat{T}^\mu_i\right)
	 + \partial_j \hat{T}^j_i = 0.
	 \label{cons-gauge}
       \end{eqnarray}

 \item $\hat{\nabla}_A \hat{T}^{A\mu} = 0$
       \begin{eqnarray}
	\frac{1}{\sqrt{h}}\nabla_\nu \left(\sqrt{h}\hat{T}^{\mu\nu}\right)
	 + g^{\mu\rho}F^{(i)}_{\nu\rho} \hat{T}^\nu_i
	 + \frac{1}{2} g^{\mu\nu}\partial_\nu h^{ij} \hat{T}_{ij}
	 + \partial_i \hat{T}^{i\mu} = 0,
	 \label{cons-EM0}
       \end{eqnarray}
\end{enumerate}
 Here $\nabla_\mu$ is the covariant derivative on the reduced space-time
 and compatible with
       $g_{\mu\nu}$ i.e. $\nabla_\mu g_{\nu\rho}=0$ and 
       $F^{(i)}_{\mu\nu} = \partial_\mu A^{(i)}_\nu - \partial_\nu
       A^{(i)}_\mu$. 

Finally if we define the energy-momentum tensor on the 
reduced space-time as
      \begin{eqnarray}
	T^{\mu\nu} &\equiv& -\frac{2}{\sqrt{-\hat{g}}}
	 \frac{\delta S}{\delta g_{\mu\nu}}
	 = \hat{T}^{\mu\nu}, \qquad
	T^{\mu}_\nu \equiv g_{\nu\rho} T^{\mu\rho} 
	= \hat{T}^\mu_\nu - A^{(i)}_\nu \hat{T}^\mu_i,
       \end{eqnarray}
     Then eq. (\ref{cons-EM0}) becomes
       \begin{eqnarray}
	\frac{1}{\sqrt{-\hat{g}}}\partial_\nu 
	 \left(\sqrt{-\hat{g}}T^\nu_\mu\right)
	 - F^{(i)}_{\mu\nu} \hat{T}^\nu_i
	 - \frac{1}{2}\partial_\mu g_{\nu\rho} \hat{T}^{\nu\rho}
	 + \frac{1}{2}\partial_\mu h_{ij} \hat{T}^{ij}
	 + g_{\mu\nu} \partial_i \hat{T}^{i\nu} = 0,
	 \label{cons-EM}
       \end{eqnarray}
       or 
       \begin{eqnarray}
	\nabla_\nu T^\nu_\mu - F^{(i)}_{\mu\nu} \hat{T}^\nu_i
	 + \frac{1}{2}h^{ij}\partial_\nu h_{ij} T^\nu_\mu
	 + \frac{1}{2}\partial_\mu h_{ij} \hat{T}^{ij}
	 + g_{\mu\nu} \partial_i \hat{T}^{i\nu} = 0.
       \end{eqnarray}
\par
The above general formalism can be easily applied to our 
case. For the Kerr black hole, the gauge current and the two-dimensional
energy-momentum tensor near the horizon is given by
\begin{eqnarray}
 J^r_{(2)} &\equiv& 
  -\int d\Omega_2 \left(r^2+a^2\cos^2\theta\right) \hat{T}^r_\varphi, \\
 T^r_{t(2)} &\equiv& \int d\Omega_2 \left(r^2+a^2\cos^2\theta\right) 
  \left(\hat{T}^r_t - A_t \hat{T}^r_\varphi\right).
\end{eqnarray}
and they satisfy the conservation equations
\begin{eqnarray}
 &&\partial_r J^r_{(2)} = 0, \\
 && \partial_r T^r_{t(2)} - F_{rt}J^r_{(2)} = 0.
\end{eqnarray}
These are the equations we used in section III. of the paper.


\begin{thebibliography}{99}


\bibitem{Iso:2006wa}
  S.~Iso, H.~Umetsu and F.~Wilczek,
  Phys.\ Rev.\ Lett.\  {\bf 96}, 151302 (2006)


\bibitem{Robinson:2005pd}
S.~P.~Robinson and F.~Wilczek,
Phys.\ Rev.\ Lett.\  {\bf 95}, 011303 (2005)


\bibitem{Hawking:sw}
S.~Hawking,
Commun.\ Math.\ Phys.\  {\bf 43}, 199 (1975).

\bibitem{Hawking:rv}
S.~Hawking,
Nature (London) {\bf 248}, 30 (1974).


\bibitem{Parikh:1999mf}
M.~Parikh and F.~Wilczek,
Phys.\ Rev.\ Lett.\  {\bf 85}, 5042 (2000);

\bibitem{tunneling}
E.~C.~Vagenas,
Phys.\ Lett.\ B {\bf 503}, 399 (2001);
%
E.~C.~Vagenas,
Mod.\ Phys.\ Lett.\ A {\bf 17}, 609 (2002);
%
E.~C.~Vagenas,
Phys.\ Lett.\ B {\bf 533}, 302 (2002);
%
A.~J.~M.~Medved,
Class.\ Quant.\ Grav.\  {\bf 19}, 589 (2002);
%
A.~J.~M.~Medved,
Phys.\ Rev.\ D {\bf 66}, 124009 (2002);
%
E.~C.~Vagenas,
Phys.\ Lett.\ B {\bf 559}, 65 (2003);
%
M.~Angheben, M.~Nadalini, L.~Vanzo and S.~Zerbini,
JHEP {\bf 0505}, 014 (2005);
%
A.~J.~M.~Medved and E.~C.~Vagenas,
Mod.\ Phys.\ Lett.\ A {\bf 20}, 2449 (2005);
%
A.~J.~M.~Medved and E.~C.~Vagenas,
Mod.\ Phys.\ Lett.\ A {\bf 20}, 1723 (2005);
%
J.~Y.~Zhang and Z.~Zhao,
JHEP {\bf 0510}, 055 (2005);
%
Q.~Q.~Jiang and S.~Q.~Wu,
Phys.\ Lett.\ B {\bf 635}, 151 (2006);
%
M.~Nadalini, L.~Vanzo and S.~Zerbini,
arXiv:hep-th/0511250;
%
J.~Zhang and Z.~Zhao,
arXiv:gr-qc/0512153.
%
Q.~Q.~Jiang, S.~Q.~Wu and X.~Cai,
Phys.\ Rev.\ D {\bf 73}, 064003 (2006);
%
S.~Q.~Wu and Q.~Q.~Jiang,
arXiv:hep-th/0603082.


\bibitem{Frolov:1998wf}
  V.~P.~Frolov and I.~D.~Novikov,
  {\it Black hole physics: Basic concepts and new developments}
  (Kluwer Academic, Dordrecht, Netherlands, 1998).

\bibitem{Christensen:jc}
S.~Christensen and S.~Fulling,
Phys.\ Rev.\ D {\bf 15}, 2088 (1977).

\bibitem{Unruh:db}
W.~Unruh,
Phys.\ Rev.\ D {\bf 14}, 870 (1976).

\bibitem{Bertlmann:xk}
R.~Bertlmann,
{\it Anomalies In Quantum Field Theory}
(Oxford Science Publications, Oxford, 2000).

\bibitem{Fujikawa:2004cx}
K.~Fujikawa and H.~Suzuki,
{\it Path integrals and quantum anomalies} 
(Oxford Science Publications, Oxford, 2004).

\bibitem{Bardeen:1984pm}
W.~A.~Bardeen and B.~Zumino,
Nucl.\ Phys.\ B {\bf 244}, 421 (1984).


\bibitem{Alvarez-Gaume:1983ig}
L.~Alvarez-Gaume and E.~Witten,
Nucl.\ Phys.\ {\bf B234}, 269 (1984).


\bibitem{Bertlmann:2000da}
R.~Bertlmann and E.~Kohlprath,
Ann.\ Phys.\ (N.Y.) {\bf 288}, 137 (2001).


\bibitem{wipf}
A. Wipf, in Black Holes: Theory and Observation, edited
by F.W.Hehl, C. Kiefer and R. Metzler (Springer, Berlin, 1998)

\bibitem{regularity}
V.~Mukhanov, A.~Wipf and A.~Zelnikov,
Phys.\ Lett.\ B {\bf 332}, 283 (1994);
%
W.~Kummer, H.~Liebl and D.~V.~Vassilevich,
Mod. \ Phys.\ Lett.\ A {\bf 12}, 2683 (1997);
%
S.~Nojiri and S.~D.~Odintsov,
Phys.\ Rev.\ D {\bf 57}, 2363 (1998);
%
R.~Balbinot and A.~Fabbri,
Phys.\ Rev.\ D {\bf 59}, 044031 (1999);
%
M.~Buric and V.~Radovanovic,
Class.\ Quant.\ Grav.\  {\bf 16}, 3937 (1999);
%
W.~Kummer and D.~V.~Vassilevich,
Annalen Phys.\  {\bf 8}, 801 (1999);
%
S.~Nojiri and S.~D.~Odintsov,
Int.\ J.\ Mod.\ Phys.\ A {\bf 16}, 1015 (2001);
%
R.~Balbinot, A.~Fabbri, V.~P.~Frolov, P.~Nicolini, P.~Sutton and A.~Zelnikov,
Phys.\ Rev.\ D {\bf 63}, 084029 (2001);
%
R.~Balbinot and A.~Fabbri,
Class.\ Quant.\ Grav.\  {\bf 20}, 5439 (2003);
%
S.~N.~Solodukhin.
arXiv:hep-th/0509148.


\bibitem{Leutweyler}
  H.~Leutwyler,
  Phys.\ Lett.\ B {\bf 153}, 65 (1985)
  [Erratum-ibid.\  {\bf 155B}, 469 (1985)].

\bibitem{soda}
 K. Murata and J. Soda, to appear.


\end{thebibliography}
\end{document}